Liquid-crystal display (LCD) of achromatic, mean-modulated flicker in clinical assessment and experimental studies of visual systems

Luke E Hallum[1,3]

Shaun L Cloherty[2]

1 Department of Mechanical Engineering, University of Auckland, Auckland, New Zealand

2 School of Engineering, RMIT University, Melbourne, Australia

3 Corresponding author: l.hallum@auckland.ac.nz

Number of words (excl. References): approx. 8300

Number of figures: 9

Number of supplementary figures: 1

Number of tables: 1

Acknowledgements: We thank Nic Price for comments on the draft manuscript, Steven Dakin for the loan of his Konica Minolta LS-110, and Emanuele Romanò for technical support. This work was partly supported by the University of Auckland Faculty of Engineering Research Development Fund.

No conflict of interest. No financial interests.

Abstract

Achromatic, mean-modulated flicker -- wherein luminance increments and decrements of equal magnitude are applied, over time, to a test field -- is commonly used in both clinical assessment of vision and experimental studies of visual systems. However, presenting flicker on computer-controlled displays is problematic; displays typically introduce luminance artifacts at high flicker frequency or contrast, potentially interfering with the validity of findings. Here, we present a battery of tests used to weigh the relative merits of two displays for presenting achromatic, mean-modulated flicker. These tests revealed marked differences between a new high-performance liquid-crystal display (LCD; EIZO ColorEdge CG247X) and a new consumer-grade LCD (Dell U2415b), despite displays' vendor-supplied specifications being almost identical. We measured displayed luminance using a spot meter and a linearized photodiode. We derived several measures, including spatial uniformity, the effect of viewing angle, response times, Fourier amplitude spectra, and cycle-averaged luminance. We presented paired luminance pulses to quantify the displays' nonlinear dynamics. The CG247X showed relatively good spatial uniformity (e.g., at moderate luminance, standard deviation 2.8% versus U2415b's 5.3%). Fourier transformation of nominally static test patches revealed spectra free of artifacts, with the exception of a frame response. The CG247X's rise and fall times depended on both the luminance from which, and to which, it responded, as is to be generally expected from LCDs. Despite this nonlinear behaviour, we were able to define a contrast and frequency range wherein the CG247X appeared largely artifact-free; the relationship between nominal luminance and displayed luminance was accurately modelled using a causal, linear time-invariant system. This range included contrasts up to 80%, and flicker frequencies up to 30

Hz. This battery of tests should prove useful to others conducting clinical assessment of vision and experimental studies of visual systems.

Introduction

Many clinical assessments and experimental studies of visual systems use flickering stimuli, specifically, achromatic, mean-modulated flicker wherein luminance increments and decrements of equal magnitude are applied, over time, to a test field, maintaining the field's time-averaged luminance. For example, loss of visual function associated with glaucoma -- a condition affecting 3.5% of 40 to 80 year-olds [1] wherein loss of retinal nerve fibres leads to loss of vision -- can be revealed by measuring a patient's behavioural responses to achromatic, mean-modulated flicker (so-called flicker-defined-form perimetry) [2, 3]. Examples of experimental studies involving human participants include the use of flicker to quantify temporal visual sensitivity [4], to selectively probe putative magno- and parvocellular visual pathways [5, 6], as well as to measure steady-state visual evoked potentials (reviewed by Norcia and colleagues [7]) and to determine retinotopic maps in cortex [8, 9].

Traditionally, vision researchers and clinicians have presented achromatic, mean-modulated flicker (as well as other dynamic visual stimuli) on computer-controlled cathode-ray tube (CRT) displays. These displays, although far from perfect, can be used to render a wide variety of dynamic stimuli. The behaviour of CRT displays has been discussed in detail in the clinical and experimental vision literature [10, 11, 12, 13, 14, 15, 16, 17] and, as a result, CRTs are usually employed in the clinic or laboratory in such a way that their imperfections do not interfere with the validity of findings. But CRT displays are now difficult to source, largely because of their recent, widespread replacement by liquid crystal displays (LCDs) in consumer-grade settings, forcing vision researchers and clinicians, in turn, to replace aged CRT displays with LCDs. LCDs are not without their own idiosyncratic behaviours as regards rendering dynamic stimuli.

However, comparatively little has been written about LCDs with the requirements of vision researchers and clinicians in mind. In the Discussion, we will review studies that have done so, comparing those results to our own.

The mechanism governing the display of light by LCDs is markedly different from that of CRT displays. In CRT displays, frames are rendered by an electron beam raster-scanned over a screen containing light-emitting phosphor [18]. The increase and decrease of luminescence of activated phosphor (i.e., the phosphor response) is fast (typically on the order of 1 ms; e.g., see [28]) relative to the rate of scanning (typically between 60 and 120 frames per second). On a given frame, the luminosity of a given point is controlled by modulating the beam current. From the point of view of the vision researcher or clinician, this mechanism confers many desirable behaviours to the CRT display, including temporal independence of luminosity from one frame to the next, even when rendering high-frequency flicker; because the phosphor response is fast, the luminosity at a point in one frame has little effect on the luminance at that point on the subsequent frame. However, a consequence of this mechanism is that all stimuli rendered by CRTs are contaminated with spatially structured, high-contrast flicker at the display's frame rate. CRT flicker is imperceptible under most viewing conditions because it exceeds the critical flicker-fusion frequency [4, 19, 20], however, it activates the early visual pathway in humans, non-human primates, and cats [21, 22], and, presumably, other experimental animal models.

LCDs comprise a spatial array of pixels, each pixel comprising a pair of electrodes, and polarizing filters that sandwich a column of liquid crystal (LC) material [23] (for a concise description of the LCD mechanism aimed at vision researchers and clinicians see Elze & Tanner [24]). The orientation of aligned LC molecules comprising the column determines the amount of light transmitted by the backlight unit to the observer; voltage applied between the electrode pair in turn applies torque to molecules, which in turn alters the molecules' orientation. A key observation is that these changes in orientation are sluggish relative to changes in voltage. Furthermore, changes in orientation are asymmetric: changes involving an increase in torque on molecules have different temporal dynamics to those involving a torque decrease [25]. These asymmetric dynamics are further complicated by digital response time compensation mechanisms, a feature of most modern LCDs designed to enhance video by accelerating step transitions between certain luminance levels [25, 26, 27, 24]. These LCD mechanisms confer many desirable behaviours to displays, including the independence of neighbouring pixels [28, 29]. However, it follows from these mechanisms that, in general, LCDs cannot be used to present achromatic, mean-modulated flicker because, when response time depends on both source and destination luminance, time-averaged luminance depends on flicker frequency and contrast.

Here we present a battery of tests used to weigh the relative merits of two displays. We are specifically interested in the presentation of achromatic, mean-modulated flicker, as exemplified by the studies cited above [2,3,4,5,6,7,8,9]. We used these tests to reveal differences between a new high-performance LCD (EIZO ColorEdge CG247X) and a consumer-grade LCD (Dell U2415b); these differences were marked, despite the displays' vendor-supplied specifications being almost identical. The CG247X's rise and fall times depended on both the luminance from which it responded as well as the luminance to which it responded, and, overall, were slow by comparison to the consumer-grade LCD. However, despite this nonlinear and sluggish behaviour, our tests revealed a contrast and frequency range wherein the CG247X's display of mean-modulated flicker was without artifacts, that is, the relationship between nominal and displayed luminance was linear.

Methods

*Displays and Visual stimuli.* We measured luminance on two LCDs: EIZO ColorEdge CG247X, and Dell U2415b. The former is a high-performance display; the latter is a consumer-grade display. Both displays were purchased in 2018, have received limited use, and have almost identical vendor-issued specifications: panel type (in-plane switching), size (61.1 cm), native resolution (1920-by-1200 pixels at 60 Hz), pixel pitch (0.270-by-0.270 mm), viewing angle (178 deg horizontal and vertical), response time (gray-to-gray between 8 and 10 ms), and light-emitting diode (LED) backlighting. We used the displays' factory default settings with one exception: the "brightness" of the U2415b was reduced to 35% to make the maximum displayed luminance of the two devices approximately equal (120 cd/m$^2$). For both displays, we used spatial resolution of 1920-by-1200 pixels and temporal resolution of 60 Hz. Luminance was rendered on displays using Psychtoolbox [30, 31, 32] (version 3.0.16) and MATLAB (version 9.5.0.944444, R2018b); Mathworks, Natick, Massachusetts, USA). We used a standard desktop computer running Linux (Ubuntu version 18.04) installed with an NVIDIA graphics card (QUADRO P620; Santa Clara, California, United States). We always allowed displays at least one hour to warm up prior to testing. Our preliminary measurements revealed no warm-up transient on either display; we found negligible difference between displayed luminance of a static test patch at one minute versus thirty minutes after power-on.

*Display linearization.* Displays were linearized in the standard fashion. We presented static test patches (384-by-384 pixels) at position 5 (Figure 2, inset) on a black background (luminance = 0%). Using a spot meter (LS-110; Konica Minolta, Tokyo, Japan) we measured displayed luminance at 11 nominal luminances (spanning the range from black to white): 0, 10, 20 … 100%. We then fit a model of displayed luminance:

$$f(x) = ax^b$$

(Eq. 1)

where $x$ is the nominal luminance, and $a$ and $b$ are free parameters. We verified the fitted model, presenting luminances $f^{-1}(0)$, $f^{-1}(10)$, $f^{-1}(20)$ … $f^{-1}(100\%)$ and ensuring the displayed luminances formed a zero-intercept straight line (e.g., see Figure 1).

**Figure 1 caption**: Example of display linearization. Open (closed) symbols show the U2415b's response before (after) linearization. Red symbols show model residuals.

*Photodiode device and Electrical recordings.* We used a device comprising a PN photodiode (BPW21R; Vishay Intertechnology, Malvern, Pennsylvania, United States) designed for applications requiring high-precision and linearity, and a custom electronic circuit (Figure S1A). The full-width at half-sensitivity of this photodiode is 105 deg, which, at a distance of 1 cm from the display, encompasses 96.3 pixels. We verified the linearity of the device by measuring the luminance of 11 static test patches (luminances = 0, 10, 20 … 100%; see *Display linearization*) with both the photodiode device and the LS-110 spot meter (Figure S1B). We also tested the photodiode device using an LED at a fixed distance of 1 cm. We used a function generator (InfiniiVision DSO-X 2002A, firmware version: 02.50.2019022736; Keysight Technologies, Colorado Springs, CO, USA) to apply square voltage waveforms (50% duty cycle) to the LED at frequencies from 1 to 32 Hz (at one-octave increments). At these frequencies, the photodiode device's response to the LED was square with no measurable change in response amplitude with frequency. We also applied a wider range of frequencies from 1 Hz to 1 MHz (at half-decade increments). The photodiode device's response was attenuated at frequencies greater than 100 kHz (Figure S1C). Photodiode device output was visualized online using our digital oscilloscope. We also used the oscilloscope to acquire electrical recordings, and to derive response times, cycle-averaged luminance, and r.m.s. luminance. This oscilloscope enables high sampling rates (up to 2 GHz); we used different sampling rates in different experiments ranging from 10 kHz to 100 kHz.

*Spatial and temporal uniformity.* To quantify the spatial and temporal uniformity of the displays, we presented static test patches (384-by-384 pixels) at nine positions (Figure 2, inset). At each position, we presented low-, medium-, and high-luminance patches (luminance = 0, 50, and 100%, respectively), measuring luminance with the LS-110 spot meter which we held by hand fronto-parallel to the patch. Furthermore, at position 5, we presented static test patches at 11 different luminances (0, 10, 20 … 100%), measuring 10 one-second recordings of luminance with the photodiode device. The photodiode was positioned 1 cm from the display, fronto-parallel, using a laser-cut acrylic arm mounted on the display housing. We computed the Fourier transforms of these photodiode device recordings using MATLAB (version 9.3.0.713579, R2017b).

*Effect of viewing angle.* To estimate the effect of viewing angle on luminance, we presented large (1920-by-1200 pixels), static test patches, measuring luminance near the center of the display with the spot meter. In these experiments, the spot meter was mounted on a tripod at 1 m. We placed the display on a turntable; to vary the angle of incidence of the spot meter's optical axis with the display, we rotated the turntable. Thus, we measured effective luminance for azimuth = -60, -40, -20, -10, 0, 10, 20, 40, 60 deg (negative angles are clockwise rotations of the turntable when seen from above). Additionally, we swiveled the display from "landscape" to "portrait" configuration, and repeated our measurements, thus varying elevation from -60 to 60 deg. Rotating the turntable (as opposed to rotating the spot meter) enabled precise control over a wide range of azimuth and elevation. However, this approach assumes that the display is spatially uniform, i.e., that spatial variability is negligibly small with respect to the effect of viewing angle. We modeled luminance as a function of azimuth, $l(\theta)$, using a circular von Mises function, i.e.,

$$l(\theta) = \frac{\alpha e^{\kappa cos(\theta)}}{2\pi I_0(\kappa)}$$

(Eq. 2)

where $I_0()$ is a modified Bessel function of the first kind, order zero. We optimized the free parameters, $\alpha$ and $\kappa$, by minimizing the sum of the squared error between the model and data. We modeled luminance as a function of elevation in the same fashion.

*Response times.* We measured response times, that is, the time taken to transition between luminances L1 and L2. L1 and L2 took values 0, 25, 50, 75, or 100%. Stimuli were square patches (384-by-384 pixels) at position 5 (Figure 2, inset). Stimuli were sequenced as follows: black (luminance = 0%) preceded stimulus L1; L1 was presented for 10 frames (166.67 ms); L2 was presented for 10 frames immediately following L1; black followed L2. We used the standard definition of rise (fall) time between luminance L1 and L2 as the duration of the transition from 10 to 90% (90 to 10%). For example, the rise time between luminance 10% and 50% was defined as the time taken to transition from 14% to 46%.

*Achromatic, mean-modulated flicker.* We presented square-wave, achromatic mean-modulated flicker at position 5 (Figure 2, inset). The time-averaged mean luminance of flicker was nominally 50%. We computed contrast (the temporal analogue of Michelson contrast) of the nominal waveform using the equation:

$$C = 100(L_{max} - L_{min})/(L_{max} + L_{min}),$$

(Eq. 3)

where $L_{max}$ is maximum luminance and $L_{min}$ is minimum luminance. We used five flicker contrasts ranging from 20 to 100%, and flicker frequencies ranging from 0.94 Hz (flicker period = 64 frames) to 30 Hz (flicker period = 2 frames).

*Paired biphasic luminance pulses.* We presented single, biphasic luminance pulses (positive/negative; see Figure 8A) at position 5 (Figure 2, inset). The mean luminance of these pulses was 50%, and their peak-to-peak amplitude was 50%. The duration of each phase was 1 frame period (16.67 ms); the duration of each pulse was 2 frames. The first of these pulses was delayed by a fixed period relative to a stable trigger, $T_t$. The second through fifth of these pulses were identical to the first pulse, but delayed by 1 through 4 frames relative to the first pulse, respectively. We also presented paired, biphasic pulses. The first pulse of all pairs was delayed by $T_t$. We parametrically varied the offset between pulses comprising the pair, T = 0, 1, 2, 3, and 4 times the frame period. We used single-pulse responses to predict each paired-pulse response. To model the display's nonlinearities, we subtracted the displayed luminance response to a paired-pulse stimulus from the prediction.

*Model of displayed luminance.* We modeled the function transferring nominal luminance to displayed luminance as a causal, exponential decay, i.e.,

$$x(t) = s(t) * k(t)$$

(Eq. 4)

where $x()$ denotes the display luminance, $s()$ denotes the nominal luminance, a square wave with 50% duty-cycle, $k()$ denotes the display transfer function, and the asterisk denotes convolution. We modelled the display transfer function, $k()$, as

$$k(t) = e^{-t/\tau}, \ t \geq 0$$

and

$$k(t) = 0, \ t < 0$$

(Eq. 5)

Wang & Nikolić [28] used a similar model albeit for a different purpose (namely, to estimate response times). Using our model response, $x()$, we derived the cycle-averaged luminance and root-mean-squared luminance for different amplitudes and frequencies of $s()$. We optimized the free parameter, $\tau$, to minimize the sum of the squared error between the model-derived measures and those derived from photodiode recordings (see *Photodiode device and Electrical recordings*).

Results

Spatial uniformity of displayed luminance can vary widely between different makes and models of LCD, the major determinant of uniformity being the backlight scheme [33] (some older LCDs allowed VGA input and relied on built-in analog-to-digital conversion, also a potential source of noise). Two commonplace schemes are, first, direct backlighting, wherein a spatial array of light-emitting diodes (LEDs) and a diffuser screen sit behind the liquid crystal panel, and, second, edge illumination, wherein light emitted by a linear array of diodes at one of the display's edges is spatially distributed via lightguide. We quantified the spatial uniformity of the CG247X by presenting low-, medium-, and high-luminance static test patches at nine display positions (Figure 2, inset) and using the LS-110 spot meter to measure the luminance of each patch. At each luminance tested, we calculated the grand average over all display positions, and divisively normalized measurements by that average. As illustrated in Figure 2, at medium- and high-luminance, the CG247X showed greater spatial uniformity than our consumer-grade LCD (Dell U2415b): for the CG247X, spatial variation was 5.1% at medium and 3.5% at high luminance, whereas for the U2415b, variation was 8.1% at medium and 8.5% at high luminance. The uniformity of the two displays was comparable at low luminance (CG247X, 27% versus U2415b, 17%). Prior to normalization, there were, as expected, marked differences between low-, medium-, and high-luminance measurements. For example, at display position 5 (Figure 2, inset) on the CG247X, low-luminance measurements ranged from 0.07 to 0.10 $cd/m^2$, medium-luminance measurements ranged from 57.70 to 57.93 $cd/m^2$, and high-luminance measurements ranged from 113.9 to 114.2 $cd/m^2$ (Table 1). We also quantified spatial surround effects; using a tripod at 1 m, we measured displayed luminance at position 5 comparing large (1920-by-1200 pixels) and small (384-by-384 pixels) 100%-luminance patches. For CG247X, the mean of 10 large-patch measurements was 0.56 cd/m$^2$ greater than that of 10 small-patch measurements (two-sample t-test, $p < 0.01$), i.e., an increase of 0.50%. For the U2415b, the increase was 0.71 cd/m$^2$, i.e., 0.67% (two-sample t-test, $p < 0.01$).

**Figure 2 caption**: Spatial uniformity of static test patches. We presented low-, medium-, and high-luminance (0, 50, 100% luminance, respectively) static test patches at nine display positions (inset), measuring luminance with the LS-110 spot meter. At each luminance tested, we calculated the grand average, and divisively normalized measurements by that average. Each symbol shows the geometric average of six normalized measurements, and error bars, where not obscured by plot symbols, show the standard error of the geometric mean. At medium- and high-luminance (grey, white symbols, respectively), the CG247X (**A**) showed greater spatial uniformity than the U2415b (**B**). Uniformity was comparable at low luminance (black symbols). The dashed square indicates the size of a test patch.

**Table 1**. Spatial uniformity of static test patches at low-, medium-, and high-luminance. Unit of measurement is cd/m$^2$.

[Table 1 here]

In-plane switching (IPS) LCDs, like our CG247X and U2415b, enable larger viewing angles than older LCD technology (e.g., twisted-nematic displays) [23]. To do so, IPS displays interdigitate electrodes (see *Introduction*) on a single substrate interposed between the backlight and the LC material [23]. For the displays we tested, vendor-issued specifications state a viewing angle of 178 deg, however, in the absence of further details, that derived measure is difficult to assimilate. We measured displayed luminance as a function of viewing angle over a range of azimuth and elevation (±60 deg). We fit a circular von Mises function (*Methods*) to measurements, determining the viewing angle that reduced effective luminance to 90% (the full-width at 90%-maximum, FW90M). As illustrated in Figure 3, the CG247X and U2415b performed comparably in this regard. For the CG247X, the FW90M was 28.6 deg (fitted parameters: $\alpha = 1.45$, $\kappa = 3.37$) and 32.6 deg ($\alpha = 1.65$, $\kappa = 2.62$) for azimuth and elevation, respectively. For the U2415b, the FW90M was 31.2 deg ($\alpha = 1.60$, $\kappa = 2.85$) and 31.0 deg ($\alpha = 1.55$, $\kappa = 2.90$) for azimuth and elevation, respectively. At high-luminance we made a reduced set of measurements, assuming rotational symmetry, varying azimuth or elevation from 0 to 60 deg. These additional measurements yielded similar FW90M estimates. This descriptive model can be used to select a viewing distance with tolerable attenuation due to viewing angle. For example, if the CG247X is viewed from 1 m, a stimulus presented at the top of the display's vertical meridian (i.e., elevation = 9.2 deg) would, due to viewing angle, undergo luminance attenuation by a factor of 0.97.

> **Figure 3 caption**: Polar plots, showing effective luminance as a function of viewing angle. We presented a large, static test patch, measuring luminance with the LS-110 spot meter near the display's center. We used a turntable to rotate the display (*Methods*), measuring the effective luminance as a function of azimuth and elevation. Each symbol shows the average of six measurements, normalized to peak. Error bars, showing the full range of measurements, were in all cases obscured by plot symbols. Solid curves show fitted circular von Mises functions (*Methods*) used to estimate full-width at 90%-maximum.

A common misconception among vision researchers and clinicians is that LCDs do not flicker (i.e., that LCDs are temporally uniform). In fact, there are two major sources of flicker that can affect a LCD: first, backlight flicker which usually occurs at temporal frequencies (e.g., 1000 Hz) well beyond the critical flicker fusion frequency (e.g., Elze & Tanner [24], and Ghodrati, Morris, & Price [34]), and, second, the so-called frame response which occurs at the refresh rate of the display (here, 60 Hz) [35, 23]. Frame responses are largely attributable to an LCD's inversion scheme: a feature of modern displays wherein the polarity of the video signal voltage applied to the liquid crystal material is inverted from one video frame to the next. This inversion minimises long-term degradation, or aging, of the display by minimizing the DC voltage across the liquid crystal elements. Frame inversion schemes typically have fine spatial structure, on the scale of individual pixels, making them mostly imperceptible (e.g., dot inversion schemes [35]). We quantified the temporal uniformity of the CG247X by presenting (nominally) static test patches at display position 5 (Figure 2, inset) and using the linearized photodiode device to measure displayed luminance over time. At each of 11 luminances (0, 10, 20 … 100%) we made 10 one-second recordings, averaging the Fourier amplitude spectra of those 10 recordings. Figure 4 shows the average spectrum at each luminance. The spectra of the CG247X revealed a frame response comprising a 60 Hz component as well as harmonic components at integer multiples of 60 Hz. The response at 60 Hz varied non-monotonically in amplitude with the luminance of the static test patch, peaking at a luminance of 50%. However, the CG247X appeared free of backlight modulations. This absence of modulations free us of the consequences of said modulations (often desynchronized with the frame refresh signal) on increment/decrement transitions between luminances (see Fig. 5 in [24]). The spectra of our consumer-grade LCD also revealed a frame response, as well as 1.2 kHz flicker, likely associated with the back light. This latter temporal nonuniformity increased linearly with the luminance of the static test patch.

For each display, we verified that the frame response was optical and not related to any radiated electromagnetic noise: We used the oscilloscope to visualize the Fourier amplitude spectrum online. We then interposed opaque cardboard between the photodiode and display which caused the disappearance of the frame response. For the U2415b, we similarly verified that the 1.2 kHz response was optical.

> **Figure 4 caption**: Temporal uniformity of nominally static test patches. We presented nominally static test patches at display position 5 (Figure 2, inset), measuring luminance with a linearized photodiode device. At each luminance (0, 10, 20 … 100%) we made ten 1-second recordings, deriving the Fourier amplitude spectrum for each. Each spectrum illustrated is the average of 10 spectra. For each display, we normalized spectra such that 1000 corresponds to the DC component at 50% luminance; therefore, a value of 5.0 corresponds to approximately 0.15 cd/m$^2$. The spectra of the CG247X (upper) revealed a frame response, comprising a 60 Hz component and harmonic components at integer multiples of 60 Hz. This frame response varied non-monotonically in amplitude with the luminance of the static test patch, peaking between 40 and 50% luminance. The spectra of the U2415b (lower) also revealed a frame response, as well as 1.2 kHz flicker, the amplitude of which increased linearly with the luminance of the static test patch (amplitudes above 5.0 are not shown, arrowheads). For the U2415b, mains noise (50 Hz) was apparent at high-luminance. lum., luminance.

In general, LCD response times -- the duration of the rise or fall of a step from one luminance level to another -- vary as a function of both step source and destination luminance. This nonlinear behaviour is owing largely to mechanisms of response time compensation (RTC) (e.g., the work of McCartney [25]), a feature of many modern LCDs designed to enhance video. RTC mechanisms speed luminance transitions by transiently altering the voltage applied to the liquid crystal associated with individual pixels (e.g., Figure 1 in [27]; Figure 5 in [24]). We measured the CG247X's response times by presenting luminance steps -- both increments and decrements -- to the linearized photodiode device. Step source and destination took values 0, 25, 50, 75, or 100%. As illustrated in Figure 5, response times varied as a function of both luminance step source and destination. For example, stepping from 0% luminance to 25% luminance took 24.5 ms, stepping from 75% to 100% took 12.9 ms, and stepping from 25% to 0% took 8.1 ms. All of these steps are the same height, but response times differ markedly. Overall, the response times of our consumer-grade LCD were less than the CG247X response times. However, as we will illustrate below, faster is not better; although RTC mechanisms reduced the response times of our consumer-grade LCD, they contaminated displayed luminance with overshoot and undershoot artifacts which are problematic for many applications in clinical and experimental vision research, including the presentation of mean-modulated flicker. RTC mechanisms lower "black-white-black" and "grey-to-grey" response times, which are used to promote displays to the gaming community and other consumer markets.

> **Figure 5 caption**: Response times. (**A**) CG247X response times. The leftmost gray box (labelled "0%") encompasses four points showing mean response times for transitions from source luminance = 0% to destination luminances = 25, 50, 75, and 100% (x axis). These rise times (upward triangles) decreased with increasing destination luminance. The gray box labelled "25%" shows mean response times of transitions from source luminance = 25% to destination luminances = 0, 50, 75, and 100%. The fall time (downward triangle), from 25% to 0% luminance, was less than the rise times. Overall, response times varied as a function of both source and destination luminance, as is generally expected of LCDs. We made 10 measurements at each source/destination luminance pair; error bars, where not obscured by symbols, mark the full range (from minimum to maximum) of these 10 measurements. (**B**) U2415b response times. Graphical conventions are as in A. Overall, U2415b response times were less than CG247X response times.

At the outset of this study, we made preliminary measurements similar to those illustrated in Figure 5. We noticed that rise and fall times straddling 50% luminance were approximately equal (e.g., rise time from 25% to 75% = 16.3 ms; fall time from 75% to 25% = 17.1 ms) which led us to wonder whether the CG247X could be used to display achromatic, mean-modulated flicker without the introduction of unworkable artifacts. To better determine the CG247X's potential suitability for presenting mean-modulated flicker, and its susceptibility, or otherwise, to overshoot and undershoot artifacts typical of LCDs implementing RTC mechanisms, we presented mean-modulated flicker on both the CG247X and our consumer-grade display, using the linearized photodiode device to measure luminance over time. We used a flicker period of 20 frames (333.3 ms), and contrast ranging from 20 to 100%. As illustrated in Figure 6, the consumer-grade display's luminance traces revealed overshoot and undershoot artifacts symptomatic of RTC. The CG247X's luminance traces, however, appeared free of RTC artifacts. We used these traces to estimate response times specific to mean-modulated flicker, illustrated in Figure 7. Overall, CG247X rise and fall times were greater than those of our consumer-grade LCD. However, with the exception of 100% contrast, CG247X rise and fall times were approximately equal, indicating its potential suitability for presenting mean-modulated flicker.

**Figure 6 caption**: Example responses of the CG247X (upper) and the U2415b (lower) to mean-modulated flicker. Flicker period = 20 frames (333.3 ms), and contrast = 20 to 100% in increments of 20 as marked. At 40% contrast, the arrowheads show examples of luminance step source and destination as used in the computation of response times (Fig. 7). For each display, we normalized traces to the luminance step destination at 100% contrast. For the U2415b, over- and undershoot are readily apparent at low and moderate contrast. The CG247X, however, shows exponential rise and fall, regardless of contrast.

**Figure 7 caption**: Response times of mean-modulated flicker. Overall, CG247X (**A**) rise (upward triangles) and fall (downward triangles) times were greater than U2415b (**B**) rise and fall times. With the exception of 100% contrast, CG247X rise and fall times were approximately equal, indicating its potential suitability for presenting mean-modulated flicker. Each symbol represents the mean of 10 measurements. Error bars, where not obscured by symbols, mark the full range (minimum to maximum) of the 10 measurements.

To further determine whether the CG247X could be used to display achromatic, mean-modulated flicker without the introduction of unworkable artifacts, we presented flicker at frequencies ranging from 0.94 to 30 Hz and contrasts ranging from 20 to 100%. We used recorded traces (similar to those in Figure 6) to derive cycle-averaged luminance. In Figure 8, we illustrate how cycle-averaged luminance was approximately constant for all flicker frequencies, and for contrasts up to 80%. At 100% contrast, cycle-averaged luminance decreased with flicker frequency, indicating that, at full contrast, the monitor is not suitable for presenting mean-modulated flicker. Cycle-averaged luminance recorded from our consumer-grade LCD (Dell U2415b) varied as a function of flicker frequency at all contrasts tested; this variation is problematic for presenting achromatic, mean-modulated flicker. We also used CG247X traces to derive cycle-averaged r.m.s. luminance. In Figure 8, we illustrate how cycle-averaged r.m.s. luminance decreased with flicker frequency, indicative of loss of contrast. The consumer-grade LCD was affected by both changes in cycle-averaged luminance and loss of contrast.

> **Figure 8 caption**: Cycle-averaged luminance and root-mean-squared luminance of mean-modulated flicker. We presented mean-modulated flicker at a range of flicker frequencies (0.94 to 30 Hz) and contrasts (20 to 100%). We used waveforms (e.g., Figure 6) recorded from the CG247X (**A**) to derive cycle-averaged luminance; we divisively normalized that derived measure using the cycle-averaged luminance of a "reference" waveform, that is, the response to contrast = 20% and flicker frequency = 0.94 Hz. This relatively low-contrast, low-frequency waveform was chosen as reference because it should be easily realized by both displays. For clarity, cycle-averaged responses for contrast = 40, 60, 80, and 100% are offset by -0.1, -0.2, -0.3, and -0.4 log units, respectively (arrowheads). As shown, cycle-averaged luminance was approximately constant for contrast = 20 to 80% at all flicker frequencies tested (0.94 to 30 Hz). At contrast = 100%, cycle-averaged luminance decreased with flicker frequency. Cycle-averaged luminance recorded from the consumer-grade U2415b (**B**) increased with flicker frequency at all contrasts tested. Graphical conventions are as in A. We used waveforms recorded from the CG247X (**C**) to derive cycle-averaged r.m.s. luminance; we divisively normalized that derived measure using cycle-averaged r.m.s. luminance of the reference waveform (20%, 0.94 Hz). As shown, at all contrasts tested (20 to 100%), cycle-averaged r.m.s. luminance decreased with flicker frequency, indicative of a loss of effective contrast. Cycle-averaged r.m.s. luminance recorded from the U2415b (**D**) revealed both increases and decreases to effective contrast with flicker frequency. Each symbol is the average of 10 measurements. (None of the data in panels C and D is offset.) We modeled cycle-average luminance and r.m.s. luminance on the CG247X as a causal exponential decay (Methods). This model comprised one free parameter, $\tau$. For the illustrated fit (blue), $\tau$ = 6.6 ms. The red symbols in panel C (slightly offset rightward for clarity) show the result of a validation experiment (see *Discussion*), wherein we used our model to stabilize cycle-averaged r.m.s. luminance across flicker frequency. Each red symbol is the average of six measurements.

Taken together, Figure 8, and the traces used to derive the measures plotted there, indicated a simple relationship between nominal and displayed luminance on the CG247X, namely, that the latter was, simply, a low-pass-filtered version of the former. To test this hypothesis, we modeled the function transferring nominal luminance to displayed luminance as a causal, exponential decay (Methods). We optimized the single free parameter in this model, the time constant of the exponential decay ($\tau$), by minimizing the sum of the squared error between the model-derived cycle-averaged mean luminance and cycle-averaged r.m.s. luminance, and those derived from the photodiode traces. For the CG247X, the fit is illustrated in Figure 8 (blue). There, the fitted parameter, $\tau$, was 6.6 ms. To assess the fit to cycle-averaged luminance, we computed the root-mean-square error (RMSE) separately at each flicker contrast. For the CG247X, the RMSE was negligibly small for contrasts from 20 to 80% (ranging from 6.0e-4 to 6.3e-3 normalized units). At 100% contrast, RMSE was highest at 0.093. This simple model was a poor fit to the U2415b, not illustrated in Figure 8. For the U2415b, RMSEs were high, ranging from 0.04 at 20% contrast to 0.15 at 60% contrast. To assess the fit to cycle-averaged r.m.s. luminance, we calculated the square of Pearson's correlation coefficient, $R^2$, separately at each flicker contrast. For the CG247X, $R^2$ was high, ranging from 0.9965 to 0.9999. As expected, the same calculation for the U2415b was consistent with a poor fit; at its worst, $R^2 = 0.03$.

To quantify the nonlinearities associated with high-contrast, mean-modulated flicker, and to quantify temporal dependence between frames, we used a paired-pulse paradigm [36, 37]. We presented paired biphasic luminance pulses at position 5 (Figure 2, inset), systematically varying the inter-pulse interval, T (Methods). We used the measured responses to individual pulses to predict paired-pulse responses, and to model the display's nonlinearities we subtracted each paired-pulse response from its prediction. Figure 9 shows the nonlinear behaviour of the CG247X and, for comparison, that of our consumer-grade LCD. In our CG247X, a nonlinear mechanism appeared to speed the transition between white and black (100% and 0% luminance, respectively; leftmost upper panel in Figure 9B). When paired pulses were separated by 16.67 ms or more (the three rightmost upper panels in Figure 9B where predicted and displayed luminance are approximately equal), the CG247X behaved linearly, that is, we saw no evidence of temporal dependence between frames. In our consumer-grade LCD, a nonlinear mechanism appeared to attenuate the transition to white (100% luminance; leftmost lower panel in Figure 9B). This attenuation reconciles with Figure 6 (lower), which shows marked overshoot at moderate contrast (e.g., 60% contrast, middlemost panel of Figure 6), but a near absence of overshoot at high-contrast (rightmost panel of Figure 6). Compared to the CG247X, the U2415b's nonlinearities were large in magnitude and long-lasting. Paired pulses separated by as much as 33.33 ms (the third lower panel in Figure 9B, where predicted and displayed luminance are unequal) evoked nonlinear behaviour in the U2415b, that is, we saw clear evidence of temporal dependence between frames.

**Figure 9 caption**: Paired luminance pulses revealed nonlinearities in displayed luminance. (**A**) Illustration of the paired-pulse paradigm. We presented a single biphasic luminance pulse (e.g., left panel), parametrically varying its latency relative to a trigger (cf. left and middle panels). We then presented a pair of biphasic luminance pulses (right panel), parametrically varying the offset between pulses comprising the pair, T = 0, 1, 2, and 3 times the frame period (frame period = 16.67 ms). Single-pulse responses can be used to predict the paired-pulse response; differences between this prediction and the displayed luminance model the display's nonlinearities. (**B**) Nonlinear behaviour of the CG247X (upper). The four panels show responses to paired pulses with various offsets, T; we normalized responses (0, 0.5 and 1 corresponded to 0, 50 and 100% luminance, respectively) and then subtracted the baseline. For each offset, the predicted displayed luminance derived from single-pulse responses is shown in blue, and the measured displayed luminance in response to paired pulses is shown in black. The measured responses are an average of 16 recordings. The difference, that is, the nonlinearity, is shown in red. For the CG247X, superposition (T = 0 ms) of pulses evoked a nonlinearity which accelerated the transition from 100% luminance to 0% luminance. There was negligible nonlinearity of displayed lumiance for T >= 16.67 ms. Compared to the CG247X's nonlinearity, the U2415b's nonlinearity (lower panels) was large in magnitude and long-lasting, affecting subsequent frames (to T = 33.33 ms). Graphical conventions are as in B.

Discussion

We have presented a battery of tests used to weigh the relative merits of two displays. We are specifically interested in the presentation of achromatic, mean-modulated flicker, as exemplified by the studies cited in the Introduction [2,3,4,5,6,7,8,9]. We recorded luminance waveforms from two LCDs, and derived several measures, including spatial uniformity, response times, Fourier amplitude spectra, cycle-averaged luminance, and root-mean-squared luminance. We presented paired luminance pulses to quantify the displays' nonlinear dynamics at high-contrast and high frequency. We find the EIZO ColorEdge CG247X is suitable for displaying mean-modulated flicker within a range of contrasts (<= 80%) and frequencies (<= 30 Hz). Unlike CRT displays, relatively little has been written in the clinical and experimental vision literature about LCDs. Here, we review detailed studies of LCD behaviour, discussing their findings with respect to ours.

*Relationship to Previous Studies*

Wang & Nikolić [28] tested a Samsung 2233RZ LCD, comparing its performance to a consumer-grade LCD, as well as a CRT display. They estimated spatial uniformity using measurement procedures similar to ours, reporting spatial variability of approximately 15% at high-luminance. Spatial variability of their CRT display (again, at high-luminance) was greater than 20%. Our CG247X, which varied by only 3.5% at high-luminance (Fig. 2), was more spatially uniform than all of their displays. Ghodrati, Morris, & Price [34] characterized the spatial uniformity of several LCDs, including two displays marketed specifically to vision researchers and clinicians (ViewPixx by VPixx Technologies Inc., Canada; and, Display++ by Cambridge

Research Systems, UK). They too used procedures like ours, but quantified uniformity by deriving Michelson contrast, $C_M = (L_{max} - L_{min})/(L_{max} + L_{min})$, where $L_{max}$ is maximum measured luminance (regardless of position 1 through 9) and $L_{min}$ is minimum measured luminance (again, regardless of position). For comparison, we derived Michelson contrast for our CG247X at low-, medium-, and high-luminance: 0.16, 0.05, and 0.03, respectively. For our consumer-grade U2415b, Michelson contrast was 0.12, 0.07, and 0.08, respectively. That is, our displays outperformed all those tested by Ghodrati, Morris, & Price (including their CRT display, for which Michelson contrast ranged from 0.096 at high-luminance to 0.76 at low-luminance), with the exception of their ViewPix, which showed very low Michelson contrast (0.023) at high-luminance. These results, taken together with LCDs' ability to render fine horizontal, oblique, and vertical gratings with little contrast attenuation [28, 29], confirm that overall the spatial uniformity of LCDs is usually superior to that of CRT displays.

Using high-luminance pulses, Wang & Nikolić [28] measured the reliability of their displays, reporting errors of less than 0.04%. Although their results are difficult to interpret because "error" was not clearly defined, they are likely to be comparable to ours: For flicker contrast at 100%, and flicker frequency at 0.94 Hz, we used 10 measurements of cycle-averaged mean-luminance, and 10 measurements of cycle-averaged r.m.s. luminance (Fig. 8), to compute coefficients of variation (for comparison, we express those coefficients as percentages). For our CG247X, we found coefficients of variation of 0.03% and 0.01% for mean-luminance and r.m.s. luminance, respectively. For our U2415b, we found coefficients of 0.01% and 0.01%, respectively. At high-contrast (100%) and high-flicker-frequency (30 Hz), for our CG247X, we found coefficients of 0.64% and 0.48% for mean-luminance and r.m.s. luminance, respectively, and for our U2415b, we found coefficients of 0.33% and 0.24%, respectively. At moderate and low frequencies (< 30 Hz), and for all contrasts we tested (20 to 100%), we found little difference between our two displays in terms of coefficient of variation of mean-luminance or r.m.s. luminance. Wang & Nikolić also reported temporal dependencies for both their 2233RZ and their CRT display, that is, displayed luminance at frame $n$ depended "to a degree on the luminance of the preceding frame [frame $n-1$]" (see their Fig. 4). We used our response time data (Fig. 5) to test our displays' temporal dependence, finding none for either our CG247X nor our consumer-grade LCD. In other words, at 3 Hz (stimulus period = 20 frames, 333.3 ms), when analyzing luminance steps, step source had no apparent effect on step destination. However, as we discuss below, our use of paired pulses revealed high-contrast, high-flicker-frequency temporal dependence between frames on our consumer-grade LCD.

Wang & Nikolić [28] measured response times (black-to-white and white-to-black, i.e., 100% contrast) of their displays by fitting exponential functions to rise and fall step responses separately. For their 2233RZ, they found fall time to be greater than rise time by a factor of 1.15. The most straightforward comparison between their results and ours is as follows: at 100% contrast, our CG247X's rise time was greater than its fall time by a factor of 1.72 (Fig. 7); therefore, the CG247X is not suitable for presenting mean-modulated flicker at 100% contrast. Matsumoto and colleagues [38] noted that approximately equal rise and fall times are key LCD behaviour for presenting flickering stimuli. At 80% contrast, we found a much smaller discrepancy between fall time and rise time: the former was greater than the latter by a factor of only 1.05. For the consumer-grade LCD tested by Wang & Nikolić, fall time was greater than rise time by a factor of 2.35. Both our CG247X and our consumer-grade LCD outperformed their consumer-grade LCD.

Our response time means (Fig. 7) were, overall, similar to those reported by other detailed characterizations of LCD behaviour [39, 24]. Our CG247X and our consumer-grade LCD mean rise (fall) times were 13.64 (17.10) ms and 8.21 (7.8) ms, respectively. The latter compares favourably to the times reported by Elze & Tanner [24]. Like Elze & Tanner, we found discrepancies between our measurements and vendor-issued responses times (CG247X: gray-to-gray = 10 ms, and black-to-white = no data; U2415b: gray-to-gray = 8 ms, and black-to-white = 19 ms). On several LCDs, Watson [39] quantified motion blur, that is, "streaking" introduced to frames when a sluggish display represents a moving object. Motion blur is a luminance artifact largely determined by response time. Watson's main measure of motion blur, Gaussian edge times (GETs), varied between approximately 7 and 22 ms across displays (see their Fig. 7). For comparison, we derived GETs for the waveforms we illustrated above in our Fig. 6. To do so, we used the "temporal step" method described by Watson, convolving those waveforms with a rectangular pulse (duration = 16.67 ms), fitting a cumulative Gaussian (Equations 1 & 2 in [39]), and recording the temporal interval between 10% and 90% points on that curve. We averaged GETs over all contrasts represented in Fig. 6 (20 to 100%), finding mean (s.d.) values comparable to Watson's: CG247X = 20.5 (0.2) ms and U2415b = 12.4 (1.8) ms. GET is a poorly performing metric where temporal responses show over- and undershoot [39]; this poor performance is reflected in the relatively large s.d. for our consumer-grade LCD (s.d. = 1.8 ms), which showed clear evidence of over- and undershoot (Fig. 6).

We used paired luminance pulses to examine the temporal dynamics of artifacts relating to high-contrast, high-frequency flicker. Ghodrati, Morris, & Price [34] also examined artifact dynamics, plotting photodiode traces for one- and two-frame-duration black-white-black pulses (see their Fig. 6). Their photodiode device was not linearized, which complicates the interpretation of their results. Nonetheless, they succeeded in demonstrating some disturbing artifacts of displayed luminance, the first and foremost being luminance blips on their Display++ (an LCD specifically marketed to vision researchers and clinicians) which appeared immediately after white-to-black transitions. None of our recordings on either of our displays revealed these trailing blips. However, our paired luminance pulses revealed nonlinearities in both our displays (Figure 9). These nonlinear mechanisms speeded transitions between extremes of luminance (CG247X), or attenuated overshoot and imparted temporal dependence between frames (U2415b).

Our spectral analysis (Fig. 4) of displayed luminance of (nominally) static test patches on both our CG247X and our consumer-grade display revealed a frame response [24, 23, 35], that is, luminance modulation associated with display refresh rate (here, 60 frames per second). For both displays, the peak-to-peak amplitude of the fundamental component (60 Hz) of this response was small -- on the order of 0.1 cd/m$^2$. However, this amplitude reflects our photodiode's spatial aperture, which was large relative to the size of single pixels; because frame inversion schemes typically have fine spatial structure on the scale of individual pixels [35], the frame response of single pixels on either of our displays was likely much larger than our measurement. Our spectral analysis confirms and extends measurements by Elze & Tanner [24], who plotted the frame response fundamental amplitude for several displays, illustrating diversity between panels from different vendors ranging from near zero (EIZO S2431W) to approximately 2% (BenQ 241W). We extend the results of Elze & Tanner, quantifying the frame response's non-monotonic dependence on luminance, peaking at mid-range luminances (Fig.

4). It is noteworthy that neither of our displays exhibited "luminance stepping", that is, saturation prior to meeting the destination luminance, which was apparent on the BenQ 241W display of Elze & Tanner, and, there, discussed in detail.

*Model Utility*

The model we developed for the CG247X (*Methods,* Equations 4 and 5), which relates nominal and displayed luminance, can be used to correct artifacts that the display introduces to achromatic, mean-modulated flicker. We illustrate by way of the following example. On the CG247X, we aimed to present flicker at a range of frequencies (from 0.94 to 30 Hz) while keeping cycle-averaged r.m.s. luminance constant, specifically, constant at that of (nominal) 80%-contrast flicker at 30 Hz. To do so, the nominal contrast of flicker was reduced at frequencies less than 30 Hz. We used our model to determine these reductions. At each frequency, we convolved the fitted exponential decay (Equation 5; $\tau$ = 6.6 ms) with a square wave (50% duty cycle), and then calculated cycle-averaged r.m.s. luminance. We then adjusted the amplitude of the square wave to yield the desired cycle-averaged r.m.s. luminance. The resulting nominal contrasts at flicker frequencies 0.94, 1.875, 3.75, 7.5, and 15 Hz were reduced from 80% to 46.5, 47.3, 48.0, 50.9, and 58.2%, respectively. In a separate experiment, we confirmed that these adjustments to nominal contrast indeed stabilized cycle-averaged r.m.s. luminance (red symbols, Fig. 8C; RMSE = 0.03). In the model of Figure 8, and the above-described example, the derived measure was cycle-averaged r.m.s. luminance. However, a similar approach could be used to fix some other derived measure of displayed luminance (e.g., cycle-averaged mean absolute luminance). For the U2415b, this straightforward approach to adjusting for artifacts is impeded by the fact that, for that display, no simple model relates nominal and displayed luminance.

*Concluding Remarks*

The above-described studies of LCD behaviour, taken together with our test results, capture a trend: for the intents and purposes of vision research, LCD performance continues to improve. Our tests revealed marked differences between two displays, despite these displays' vendor-supplied specifications being almost identical. In other words, these specifications were entirely useless in weighing the relative merits of our displays for presenting achromatic, mean-modulated flicker. One of the displays tested -- the EIZO ColorEdge CG247X -- appears largely free of artifacts when presenting flicker within a restricted range of contrasts and frequencies. We hope the present battery of tests proves useful to others developing rigorous approaches to clinical assessment of vision and experimental studies of visual systems; we intend to use it when introducing new displays to our experimental set-ups.

DISPLAY OF MEAN-MODULATED FLICKER

# Table 1

| | | CG247X | | | | | | | | |
|---:|---:|---:|---:|---:|---:|---:|---:|---:|---:|---:|
| | | Display Position [1] | | | | | | | | |
| | | 1 | 2 | 3 | 4 | 5 | 6 | 7 | 8 | 9 |
| Low luminance | mean [2] | 0.08 | 0.09 | 0.11 | 0.08 | 0.09 | 0.08 | 0.08 | 0.08 | 0.09 |
| | s.e.m. | 0.003 | 0.004 | 0.003 | 0.002 | 0.005 | 0.003 | 0.002 | 0.002 | 0.002 |
| Medium | mean | 62.0 | 59.9 | 60.2 | 59.6 | 57.8 | 59.0 | 57.9 | 56.5 | 58.0 |
| | s.e.m. | 0.018 | 0.074 | 0.138 | 0.054 | 0.031 | 0.104 | 0.038 | 0.066 | 0.042 |
| High | mean | 118.7 | 117.4 | 114.4 | 115.3 | 114.0 | 114.1 | 113.3 | 111.0 | 113.6 |
| | s.e.m. | 0.281 | 0.120 | 0.531 | 0.293 | 0.039 | 0.113 | 0.129 | 0.099 | 0.039 |

| | | U2415b | | | | | | | | |
|---:|---:|---:|---:|---:|---:|---:|---:|---:|---:|---:|
| | | Display Position | | | | | | | | |
| | | 1 | 2 | 3 | 4 | 5 | 6 | 7 | 8 | 9 |
| Low luminance | mean | 0.10 | 0.11 | 0.09 | 0.09 | 0.10 | 0.09 | 0.09 | 0.09 | 0.09 |
| | s.e.m. | 0.003 | 0.003 | 0.004 | 0.004 | 0.002 | 0.002 | 0.002 | 0.004 | 0.003 |
| Medium | mean | 54.6 | 56.2 | 53.7 | 51.3 | 53.4 | 52.1 | 49.0 | 48.8 | 48.6 |
| | s.e.m. | 0.070 | 0.147 | 0.090 | 0.124 | 0.099 | 0.024 | 0.159 | 0.150 | 0.042 |
| High | mean | 108.9 | 114.6 | 111.2 | 102.3 | 109.8 | 107.5 | 97.0 | 99.3 | 100.0 |
| | s.e.m. | 0.303 | 0.200 | 0.266 | 0.233 | 0.176 | 0.119 | 0.093 | 0.266 | 0.145 |

Table notes:
[1] See Figure 2, inset
[2] Arithmetic mean of 6 measurements
s.e.m., standard error of the mean

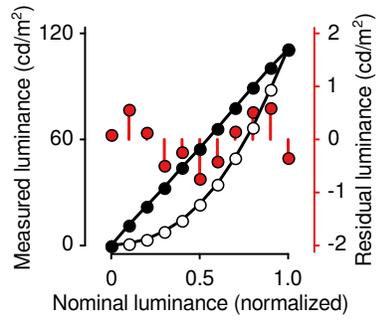

*Figure 2*

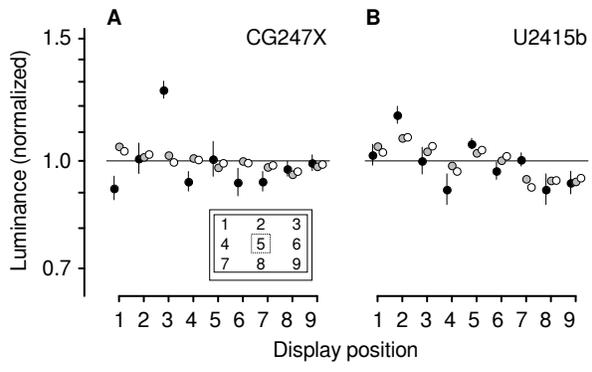

*Figure 3*

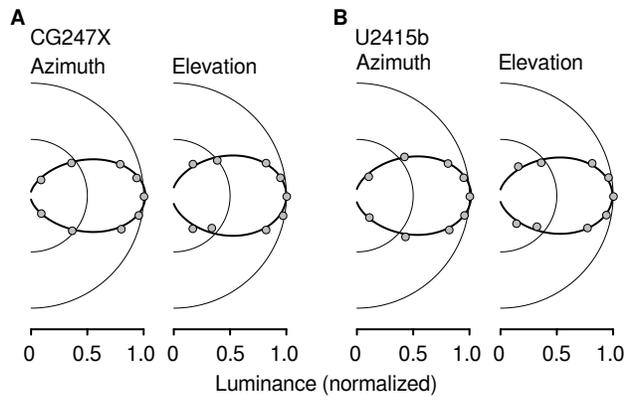

*Figure 4*

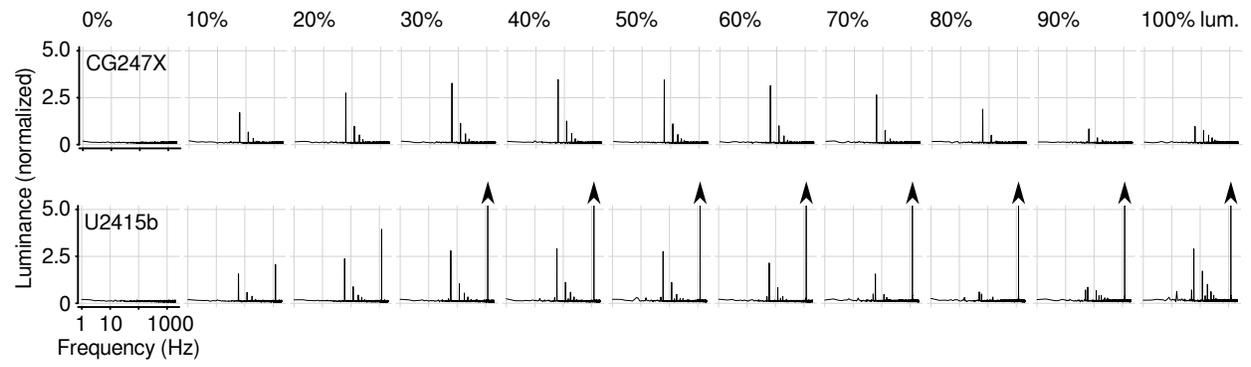

Figure 5

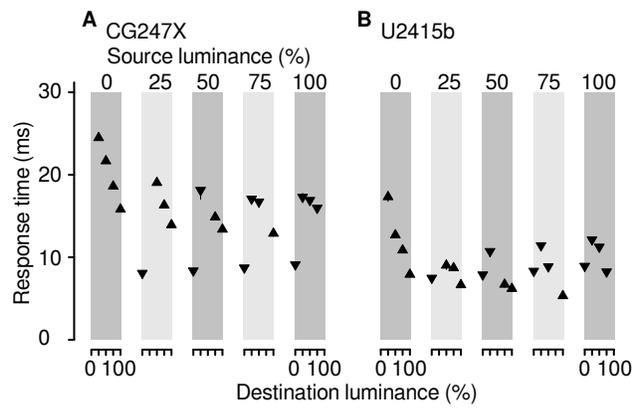

*Figure 6*

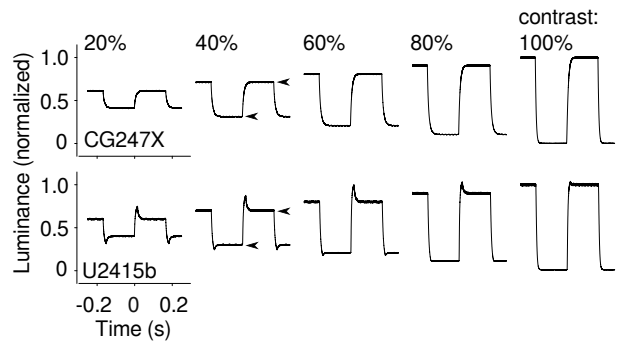

*Figure 7*

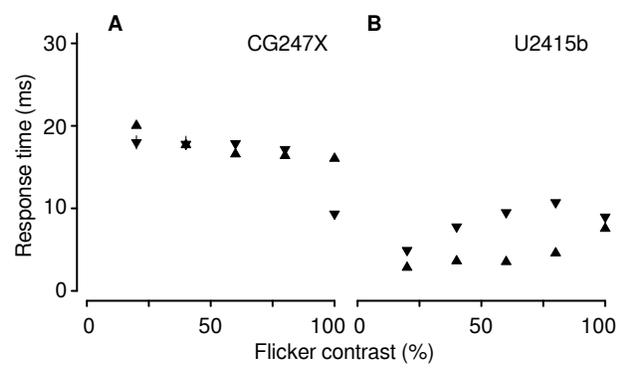

*Figure 8*

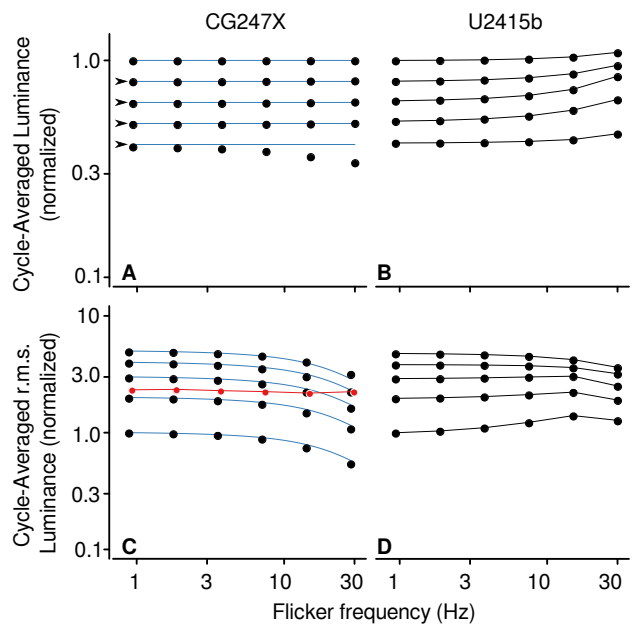

Figure 9

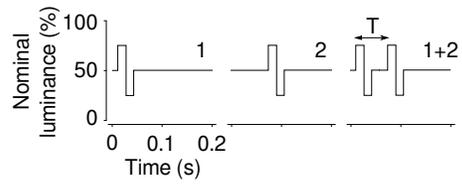

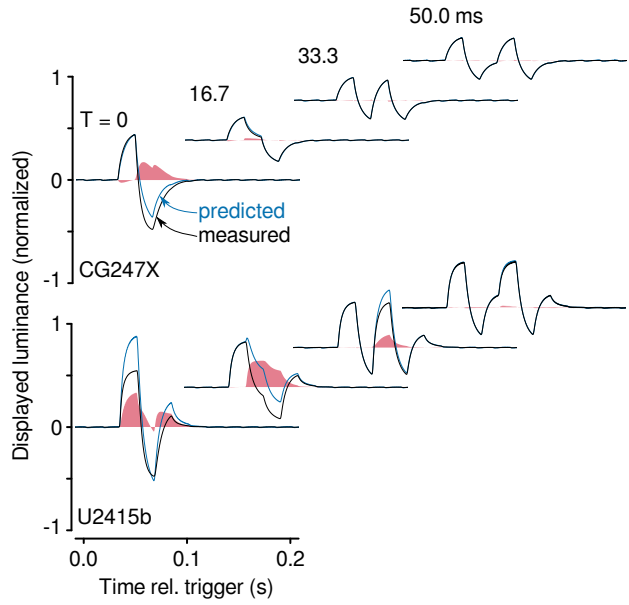

# Figure S1

**A**

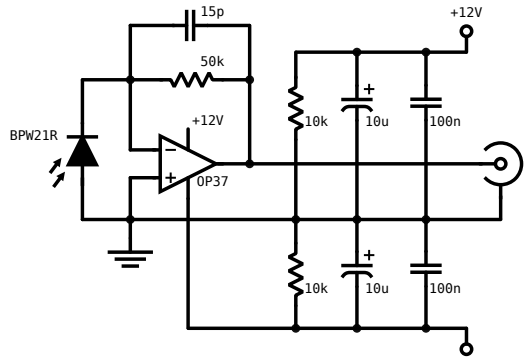

**B**

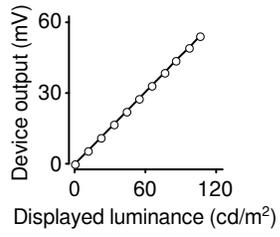

**C**

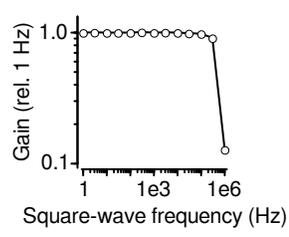